# Selective interlayer ferromagnetic coupling between the Cu spins in $YBa_2Cu_3O_{7-x}$ grown on top of $La_{0.7}Ca_{0.3}MnO_3$

S. W. Huang[1,2,3], L. Andrew Wray[1,4,5], Horng−Tay Jeng[6,7], V. T. Tra[8], J. M. Lee[9], M. C. Langner[2], J. M. Chen[9], S. Roy[1], Y. H. Chu[10], R. W. Schoenlein[2], Y.−D. Chuang[1,*], and J.−Y Lin[8,1,+]

[1]Advanced Light Source, Lawrence Berkeley National Laboratory, Berkeley, CA 94720, USA
[2]Materials Sciences Division, Lawrence Berkeley National Laboratory, Berkeley, CA 94720, USA
[3]MAX IV Laboratory, Lund University, P. O. Box 118, 22100 Lund, Sweden
[4]Department of Physics, New York University, New York, New York 10003, USA
[5]Stanford Institute for Materials and Energy Sciences, SLAC National Accelerator Laboratory, Menlo Park, CA 94025, USA
[6]Department of Physics, National Tsing Hua University, Hsinchu 30013, Taiwan
[7]Institute of Physics, Academia Sinica, Taipei 11529, Taiwan
[8]Institute of Physics, National Chiao Tung University, Hsinchu 30010, Taiwan
[9]National Synchrotron Radiation Research Center, Hsinchu 30076, Taiwan
[10]Department of Materials Science and Engineering, National Chiao Tung University, Hsinchu 30010, Taiwan
*ychuang@lbl.gov
+ago@cc.nctu.edu.tw

## ABSTRACT

Studies to date on ferromagnet/d-wave superconductor heterostructures focus mainly on the effects at or near the interfaces while the response of bulk properties to heterostructuring is overlooked. Here we use resonant soft x-ray scattering spectroscopy to reveal a novel c-axis ferromagnetic coupling between the in-plane Cu spins in $YBa_2Cu_3O_{7-x}$ (YBCO) superconductor when it is grown on top of ferromagnetic $La_{0.7}Ca_{0.3}MnO_3$ (LCMO) manganite layer. This coupling, present in both normal and superconducting states of YBCO, is sensitive to the interfacial termination such that it is only observed in bilayers with $MnO_2$ but not with $La_{0.7}Ca_{0.3}O$ interfacial termination. Such contrasting behaviors, we propose, are due to distinct energetic of CuO chain and $CuO_2$ plane at the $La_{0.7}Ca_{0.3}O$ and $MnO_2$ terminated interfaces respectively, therefore influencing the transfer of spin-polarized electrons from manganite to cuprate differently. Our findings suggest that the superconducting/ferromagnetic bilayers with proper interfacial engineering can be good candidates for searching the theorized Fulde-Ferrel-Larkin-Ovchinnikov (FFLO) state in cuprates and studying the competing quantum orders in highly correlated electron systems.

## Introduction

Ferromagnetism and d-wave superconductivity are often viewed as antagonistic orders as the spin exchange field from ferromagnetism can introduce energy difference between electrons in the spin-singlet Cooper pair. Although the Cooper pair can also be formed with electrons from Zeeman splitted Fermi surfaces, an approach that gives finite center of mass momentum to the Cooper pair and leads to a spatially modulated superconducting order parameter, such state (Fulde-FerrelLarkin-Ovchinnikov, or FFLO, state[1–3]) remains to be identified in the high temperature superconducting cuprates. The coexistence of ferromagnetism and superconductivity has been reported in some uranium-based superconductors,[4, 5] but these superconductors have p-wave pairing symmetry and will not be the right candidates for studying the competitive interactions between ferromagnetism and d-wave superconductivity.

With advanced thin-film growth technology,[6–10] heterostructures made out of superconducting cuprates and ferromagnetic manganites can serve as an ideal platform for such studies. Measurements on $YBa_2Cu_3O_{7-x}/La_{0.7}Ca_{0.3}MnO_3$ (YBCO/LCMO) heterostructures have revealed an induced ferromagnetic Cu spin moment in the interfacial $CuO_2$ plane that couples antiferromagnetically to the underlying Mn spin moment.[6] Interesting phenomena such as ferromagnetism/superconductivity proximity[11,12] and inverse proximity effect,[13] transfer of spin-polarized electrons from manganite to cuprate,[14] and the electronic orbital reconstruction at the interface[7] were also observed and proposed to account for the suppression of both ferromagnetism and d-wave superconductivity upon forming such heterostructures (Fig. 1(a)). However, studies to date on these heterostructures focus mainly on the effects at or near the interfaces while the response of bulk properties to heterostructuring is overlooked. Here we use bulk-sensitive resonant soft x-ray scattering spectroscopy (RSXS) and first principles calculations to show that beyond the interface, a novel ferromagnetic order can be established within the YBCO layer possibly even in the superconducting state. This c-axis ferromagnetic coupling between the in-plane Cu spins is subtle and can be effectively controlled by the interfacial termination.

## Results

### X-ray reflectivity of the bilayer

Due to very small mismatch between the in-plane lattice constants of YBCO and LCMO, epitaxial growth of YBCO/LCMO bilayers with atomically smooth interfaces can be achieved. As illustrated in Fig. 1(b), the crystallinity requires that $CuO_2$ plane and CuO chain to be at the cuprate/manganite interface in $MnO_2$ (left panel) and $La_{0.7}Ca_{0.3}O$ (right panel) terminated bilayers, respectively. After growth, the film quality was checked separately with synchrotron reflectivity and the result from $MnO_2$ terminated bilayer at 80 K is shown in Fig. 1(c). The measurement photon energy was 1240 eV, far away from Mn and Cu resonances. Clear intensity oscillations known as Kiessig fringes from the constructive and destructive interferences between the reflected x-rays off different interfaces confirm the high film quality. By fitting these Kiessig fringes, the thickness of each layer is determined be 15 nm (YBCO) and 7.5 nm (LCMO), in agreement with the growth conditions (see section V in Supplementary Information).

One immediately notices that the intensity of YBCO (001) structural Bragg peak, marked by the arrow in the figure, is only slightly stronger than the Kiessig fringes at this photon energy. This peak is even weaker than the (002) Bragg peak measured at larger 2θ angle. The weak (001) Bragg peak is due to the unique YBCO form factor such that the scattered x-rays from charges in two $CuO_2$ planes and one CuO chain in the unit cell interfere destructively, and is the key to allow us to identify the even weaker magnetic contribution.

### Comparison of XAS spectra and RSXS resonance profiles

The electronic structure of YBCO is changed when it is grown on top of the LCMO layer. These changes can even be seen at 300 K where the LCMO layer remains paramagnetic. Thin solid lines in Fig. 2 represent the Cu L-edge x-ray absorption (XAS) spectra from a pure YBCO film (green, top panel) and the bilayers with $MnO_2$ (red, middle panel) and $La_{0.7}Ca_{0.3}O$ (blue, bottom panel) interfacial terminations recorded in the total electron yield mode. Although the main peak at 925.75 eV, ascribed to in-plane $Cu^{2+}$ with a single $3d_{x^2-y^2}$ hole,[15] is very similar between these samples, the shoulder near 927.5 eV and the high energy feature around 928.75 eV exhibit small differences. The shoulder structure, relevant to the ligand holes in $CuO_2$ planes ($Cu^{2+}L$, see inset in Fig. 2), is suppressed in both bilayers. The suppression is consistent with their lower superconducting transition temperatures from effectively lower hole doping levels.[16] In addition, the high energy feature from $Cu^+$ shows an enhancement, implying a noticeable charge-transfer effect taking place in CuO chains.

Even though these changes in XAS spectra upon heterostructuring are subtle, they can be clearly seen in the resonance profiles. In RSXS spectroscopy, the resonance profile $|F(\vec{q},E)|^2$ of an electronic order is sensitive to the electronic states that coherently scatter the x-rays into the specific ordering wave vector $\vec{q}$ (see Methods and the schematic of experimental setup in Fig. 3(e)).[17,18] $F(\vec{q},E)$ depends on the form factor of scattering channel(s) of $i^{th}$ atom $f_i^{(\hat{\varepsilon}_i,\hat{\varepsilon}_o)}(E)$, which can be charge, spin and orbital degrees of freedom, and the spatial arrangement of these scatterers $e^{i\vec{q}\cdot\vec{r}_i}$ein the following way: $F(\vec{q},E) = \sum_i f_i^{(\hat{\varepsilon}_i,\hat{\varepsilon}_o)}(E)e^{i\vec{q}\cdot\vec{r}_i}$. Here $\vec{r}_i$ is the position vector of $i^{th}$ atom, E is the excitation photon energy and the summation is carried out over the superlattice. $f_i^{(\hat{\varepsilon}_i,\hat{\varepsilon}_o)}$ depends on the incident ($\hat{\varepsilon}_i$) and scattered ($\hat{\varepsilon}_o$) photon polarizations, and is the sum of real ($f'_i(E)$) and imaginary ($f''_i(E)$) parts that are related to each other through Kramers-Kronig relations. Because of such dependence, RSXS resonance profile often exhibits a higher degree of sensitivity to changes in local electronic structure than XAS. In the current study, we choose YBCO (001) Bragg peak ($\vec{q}$ = (001) in (HKL) unit) since the ferromagnetic coupling between magnetic unit cells along c-axis would have the same wave vector as the charge unit cell.

The resonance profiles of these samples are shown as thick solid lines in Fig. 2. For bilayers, irrespective to which interfacial termination, the resonance profiles are very different from that of the pure YBCO film. Their resonance profiles show two prominent peaks instead of one at both Cu $L_3$ and $L_2$ edges (features labelled A and B at $L_3$ edge). This double-peak structure is intrinsic and is neither caused by the self-absorption effect in the fluorescence yield measurements, nor by the presence of two types of $Cu^{2+}$ in $CuO_2$ planes. Further discussions can be found in section III in Supplementary Information. Comparison to the maximum in XAS spectra shows that feature A is shifted towards lower photon energy by ~0.5 eV. Since XAS and RSXS spectra in Fig. 2 were recorded simultaneously, this energy shift is not an experimental artifact. In fact, similar energy shift has been reported in some transition metal oxides that exhibit electronic ordering phenomena and in the case of cuprates, it was previously attributed to the subtle spatial variation in the local energetic of Cu 3d and O 2p states.[19–21] The detailed calculation of resonance profiles remains an actively developing field and will not be discussed in this paper. For example, one may attempt to first use the x-ray magnetic reflectivity (XRMR) to fit the Kiessig fringes to obtain the complex index of refraction $n(E) = 1 - \delta(E) + i\beta(E)$, which contains the information of charge and magnetic density profile over the thickness of layers, and then use the optical theorem to link $\delta(E)$ and $\beta(E)$ to $f'(E)$ and $f''(E)$ respectively.[22] However, this approach in only valid in the forward scattering geometry where Kiessig fringes are the dominant features in the spectra. In addition, XRMR cannot reveal the local spin moment. For rigorous treatment of resonant x-ray scattering using the Kramers-Heisenberg formalism, see ref.[23, 24] Here we set the measurement photon energy at 925.25 eV to focus on feature A to study the magnetic interactions in $CuO_2$ planes. At this photon energy, the 2θ angle is around 70°. Inset in Fig. 1(c) shows the exemplary q-scans at 80 K (blue) and 300 K (red) from $MnO_2$ terminated bilayer at this photon energy. As one can see, the Kiessig fringes that overwhelm the (001) Bragg peak in Fig. 1(c) become negligible. The correlation length, determined from the inverse of half-width-half-maximum of the peak, is >12 nm and is in agreement with the YBCO layer thickness. This again confirms the bulk nature of recorded RSXS signal.

**Additional magnetic component in the YBCO (001) Bragg peak**

Fig. 3 shows the main experimental findings of this paper: the temperature dependence of normalized YBCO (001) Bragg peak intensity (red markers, left axis; the intensity is normalized to 1.0 at 300 K) overlaid with the magnetization curves (blue lines, right axis) from these samples. The data points shown here are from the Lorentzian fitting of q-scan spectra (see Methods). For pure YBCO film, the temperature-independent Bragg peak intensity implies that the changes in Cu charge scattering form factor, as well as their spatial arrangement, are negligible between . 80 K and 300 K as expected (Fig. 3(a)). On the other hand, the Bragg peak intensity from $MnO_2$ terminated bilayer shows intriguing temperature dependence where two step-like increases can be correlated with characteristic temperatures in the magnetization curve: the . 190 K is the Curie temperature of LCMO layer and the . 105 K is related to the structural phase transition of STO substrate (Fig. 3(b)).[25] Since we do not expect to see changes in Cu charge scattering as temperature is lowered, the increases can be attributed to an additional order induced by the ferromagnetism in LCMO layer.

The nature of this additional order can be further investigated by looking at the Bragg peak intensity as a function of sample orientation relative to the photon polarization (see Fig. 3(e) for experimental geometry and the Methods). Under the resonance condition, the scattering intensity $|F(E)|^2$ can come from following terms that involve (unit vector) $\hat{\varepsilon}_i$, $\hat{\varepsilon}_0$ and spin moment $\hat{s}$: $\hat{\varepsilon}_i \cdot \hat{\varepsilon}_0$, $(\hat{\varepsilon}_i \times \hat{\varepsilon}_0) \cdot \hat{s}$ and $(\hat{\varepsilon}_i \cdot \hat{s})(\hat{\varepsilon}_0 \cdot \hat{s})$.[17, 18] These three terms contribute to the monopole charge scattering, circular and linear dichroism respectively. Since the incident photon energy is tuned close to $Cu^{2+}$ $L_3$ absorption edge, we only consider the dipole (E1) transition. Following the treatment by Hill & McMorrow,[17] the scattering tensor can be expressed with components that depend on the incident and scattered photon polarizations (here, for example, $\sigma_i$ and $\pi_0$ refer to incident σ and scattered π polarization respectively):

$$f^{(\hat{\varepsilon}_i, \hat{\varepsilon}_0)} \to \begin{pmatrix} f_{\sigma_i \to \sigma_0} & f_{\pi_i \to \sigma_0} \\ f_{\sigma_i \to \pi_0} & f_{\pi_i \to \pi_0} \end{pmatrix}$$

$$= F^{(0)}(E) \underbrace{\begin{pmatrix} 1 & 0 \\ 0 & \cos(2\theta) \end{pmatrix}}_{\hat{\varepsilon}_i \cdot \hat{\varepsilon}_0} + iF^{(1)}(E) \underbrace{\begin{pmatrix} 0 & z_1 \cos(\theta) + z_3 \sin(\theta) \\ z_3 \sin(\theta) - z_1 \cos(\theta) & -z_2 \sin(2\theta) \end{pmatrix}}_{(\hat{\varepsilon}_i \times \hat{\varepsilon}_0) \cdot \hat{s}}$$

$$+ F^{(2)}(E) \underbrace{\begin{pmatrix} z_2^2 & -z_2(z_1 \sin(\theta) - z_3 \cos(\theta)) \\ z_2(z_1 \sin(\theta) + z_3 \cos(\theta)) & -\cos^2(\theta)(z_1^2 \tan^2(\theta) + z_3^2) \end{pmatrix}}_{(\hat{\varepsilon}_i \cdot \hat{s})(\hat{\varepsilon}_0 \cdot \hat{s})} \quad (1)$$

$F^{(0)}(E)$, $F^{(1)}(E)$ and $F^{(2)}(E)$ are defined in ref.[17] $z_i$ are the spin moments projected along three Cartesian axes and θ is the grazing incidence angle (~35° in the current study).

Usually, the $F^{(i)}(E)$ terms do not mix with each other except that the $0^{th}$ harmonic component of $F^{(2)}(E)$ can mix with $F^{(0)}(E)$. But for ferromagnetic coupling where the magnetic unit cell coincides with the charge unit cell, the Kronecker δ conserving the wave vectors in each term becomes 1 and these three terms need to be considered all together as shown above.

In calculating the scattering intensity by taking the square of this $f^{(\hat{\varepsilon}_i, \hat{\varepsilon}_0)}$ matrix in equation (1), the cross terms $[F^{(1,2)} * F^{(0)}]$ will enhance the weak magnetic signal. Keeping the leading terms, the scattering intensity will vary with sample azimuthal angle $(\varphi)$ as following: $a + b \cdot \sin(\varphi) + c \cdot \cos^2(\varphi)$. The parameters b and c depend on the spin projection angle α and are proportional to $\left|\frac{F^{(1)}}{F^{(0)}}\right|$ and $\left|\frac{F^{(2)}}{F^{(0)}}\right|$ (see Section IV in Supplementary Information).

In Fig. 3(f), the (001) Bragg peak intensity at 80 K shows the strong $\varphi$ dependence, which can be fitted with the aforementioned functional form (blue curve in Fig. 3(f)). One should note that such $\varphi$ dependence cannot come from $Cu^{2+}$ charge scattering or the spin component normal to the $CuO_2$ plane because these two contributions do not depend on $\varphi$. On the other hand, it can come from $Cu^{2+}$ spin component in the $CuO_2$ plane. This finding implies the magnetic origin for the additional order seen in the YBCO (001) Bragg peak. The (001) wave vector further tells us that it is caused by the inter-unit-cell (c-axis) ferromagnetic coupling between the in-plane Cu spins. With the p-scattering geometry, the maxima around 0° and 180° allow us to determine that the projected in-plane component is along the Cu-O bond direction (Fig. 3(e) shows the geometry at $\varphi=0°$). This spin alignment is 45° away from the easy axis of Mott antiferromagnetism, which is along the Cu-Cu or (110) direction, but follows the easy axis of ferromagnetism in the LCMO layer and is likely the outcome of strong couplings between manganite and cuprate.[6, 29, 30]

Although the induced ferromagnetic spin moments in the interfacial $CuO_2$ plane below the Curie temperature of LCMO layer have been previously reported,[6, 26–28] our RSXS data suggests that they may have coupled ferromagnetically along c-axis throughout the YBCO layer as the diffraction peak width is limited by the layer thickness. This c-axis ferromagnetic spin coupling is established at 80 K, just above the ∼ 70 K superconducting transition temperature of YBCO under study. But the continued increase in RSXS intensity in Fig. 3(b) points to the scenario that this coupling can persist down to 30 K for its absence would put the normalized intensity value back to ∼1.0. But this coupling is subtle and can be greatly influenced by the interfacial termination. We have performed the same RSXS measurements on $La_{0.7}Ca_{0.3}O$ terminated bilayer and the results are shown in Fig. 3(c). Despite showing very similar characteristic temperatures in the magnetization curve as those in the $MnO_2$ terminated bilayer, the (001) Bragg peak intensity remains nearly temperature independent within our measurement resolution.

## DFT calculations

To investigate these contrasting behaviors, we have carried out the DFT calculations (for details, see Methods). The schematic in Fig. 4 shows the stacking order of CuO chains (black filled circles with vertical bars) and $CuO_2$ planes (red open circles with horizontal bars) along c-axis in the calculations. The labeling of Cu sites is guide for readers to associate the calculated spin moments with their spatial arrangement. The positive and negative moments refer to the Cu spins that are parallel and antiparallel to the Mn spins, respectively. The DFT calculations show that the magnitude of Cu spin moment in the $MnO_2$ terminated bilayer (Fig. 4(a)) is on the order of $\sim 0.02 \mu_B$/Cu in $CuO_2$ planes (red open circles), consistent with the XMCD measurements. However, the moment in CuO chains is negligible (black filled circles). The antiferromagnetic coupling between Mn and Cu spin moments near the interface is correctly reproduced in our DFT calculations (Cu site #1 in Fig. 4(a)). This coupling remains antiferromagnetic between the first two $CuO_2$ planes, a phenomenon that was previously predicted by the model calculations and was attributed to an anomalous screening effect.[31] Besides the first unit cell, our DFT calculations predict the ferromagnetically coupled spin moments through out the rest of YBCO layer. In contrast, for $La_{0.7}Ca_{0.3}O$ terminated bilayer, only the chain Cu right next to the interface exhibits a finite spin moment that couples ferromagnetically to the Mn spin moment (Fig. 4(b)). In that regard, the $La_{0.7}Ca_{0.3}O$ terminated bilayer is not expected to have ferromagnetic order inside the YBCO layer. Our DFT calculations have given the results that are in agreement with the RSXS data in Fig. 3(b) and 3(c).

## Discussion

We have identified a novel inter-unit-cell ferromagnetic coupling between the in-plane Cu spins in YBCO based on the clear $\varphi$ and temperature dependence of YBCO (001) Bragg peak intensity recorded at $Cu^{2+}$ resonance energy. This ferromagnetic coupling is sensitive to the interfacial termination such that we could not detect its presence in the $La_{0.7}Ca_{0.3}O$ terminated bilayer under the same experimental condition. Observation of this coupling can also be compared with previous model calculations.[31] Here, we propose that the distinct energetic of $CuO_2$ plane and CuO chain at the manganite/cuprate interface in the $MnO_2$ and $La_{0.7}Ca_{0.3}O$ terminated bilayers respectively is responsible for the contrasting behaviors.

For $MnO_2$ terminated bilayer, we propose that the double-exchange interaction in the itinerant $e_g$ bands of YBCO,

which is absent in the pristine YBCO and was omitted in the previous model calculations,[31] emerges due to the induced Cu spin moments and the influence from the ferromagnetism in poorly screened LCMO underlayer.11 For $La_{0.7}Ca_{0.3}O$ terminated bilayer, the situation is rather different. Although the chain Cu has higher affinity to attract the spin-polarized electrons from LCMO layer, as evident from its lower superconducting transition temperature, the larger induced moment (~$0.04\mu_B$/Cu) and the ferromagnetic coupling to the Mn spin moment, the lower $e_g$ electron itinerancy in quasi-1D chains plus the strong electron affinity can localize the transferred electrons to the interface and weaken the double-exchange interaction in the remaining YBCO layer (Fig. 4(b)). Furthermore, different occupancy on the respective orbitals of chain and plane Cu can also disrupt the double-exchange mediated ferromagnetic coupling.

Since previous XMCD measurements adopted specific geometries to suppress the ferromagnetic signal from the bulk layers so as to enhance the contrast from the interface region, these XMCD results are not able to substantiate or dispute our findings.[6, 26, 27, 32] Therefore, observing this c-axis ferromagnetic coupling calls for further investigation on the $MnO_2$ terminated bilayers using XMCD with geometries that emphasize the sensitivity to bulk ferromagnetic signal from YBCO layer. In addition, our DFT calculations suggest that the induced Cu spin moment can couple ferromagnetically to the Mn spin moment in the $La_{0.7}Ca_{0.3}O$ terminated bilayer. Thus performing the XMCD measurements on this type of bilayer can serve as an independent check to the theories. Our results also highlight the fact that the bulk properties of constituting layers will respond to heterostructuring, an aspect that is largely overlooked, and it has recently been shown that interesting phenomena can be manifested beyond the interfaces by applying the perturbations in the heterostructures.[33]

In conjunction with previous XMCD measurements,[6, 26, 27, 32] our results suggest that the induced ferromagnetic Cu spin moments in $CuO_2$ planes may couple ferromagnetically along the c-axis, forming the bulk ferromagnetism in YBCO. This ferromagnetism can even exist in the superconducting state. It is known that the correlated transition metal oxides can exhibit strong tendency towards electronic phase separation, which leads to other intriguing phenomena like colossal magnetoresistance,[37] but the possibility of coexistence of ferromagnetism and d-wave superconductivity (ferromagnetic d-wave superconductor) makes this type of bilayer an exciting platform for investigating novel phases associated with high temperature superconductivity. Introducing ferromagnetism to relax the asymptotic confinement that limits the carrier mobility in the underdoped regime not only serves as an alternative to induce the emergent quantum states besides the conventional hole doping approach, but can also reveal a new dimension in cuprate phase diagram. Subjecting d-wave superconductivity to strong exchange field weakens its strength and enhances the competition with other ground states such as charge checkerboard and/or stripes,[34, 35] and examining the extent of such competitions can be facilitated by using the heterostructures. Moreover, realizing the prerequisites for FFLO state implies that cuprate/manganite heterostructures may exhibit other exotic properties like non-centrosymmetric Cooper pairs and stripe-like inhomogeneity.[36] Exploring their non-trivial interplay with inherent charge inhomogeneity will expand the boundaries of d-wave superconductivity theories. To rule out the ubiquitous phase separation phenomenon[38] and unambiguously identify this ferromagnetic d-wave superconductor, spectromicroscopy with sufficient energy and spatial resolutions such as spin-resolved nano-ARPES can be the decisive tool for this task.[39]

## Methods
### Materials
$YBa_2Cu_3O_{7-x}/La_{0.7}Ca_{0.3}MnO_3$ bilayers were fabricated on top of the (100) oriented $SrTiO_3$ (STO) single crystal substrate using pulsed laser deposition (PLD) method. A KrF (λ = 248 nm) excimer laser, with 10 Hz repetition rate and 250 mJ power, was used to evaporate the targets. Before growing the bilayers, the substrate was first treated with HF-NH4F buffer solution to produce a uniform $TiO_2$ termination at the surface. In-situ reflection high energy electron diffraction (RHEED) was used to monitor the layer growth. The LCMO and YBCO layers were deposited at 700°C and 750°C, and 80 mTorr and 150 mTorr oxygen pressures respectively. After growth, these films were annealed in a 700 Torr oxygen atmosphere at 550 °C for one hour followed by slow cooling down to the room temperature to achieve full oxygenation for the YBCO layer. To produce the $MnO_2$ termination at the manganite/cuprate interface, we started with $TiO_2$ terminated STO substrate and deposited the LCMO layer directly on top of it, then followed by the deposition of YBCO layer. To switch to the $La_{0.7}Ca_{0.3}O$ termination at the interface, a $SrRuO_3$ buffer layer (SRO, 1.5 u.c.) was deposited between the STO and LCMO layer. The schematic crystal structures shown in Fig. 1(b) were confirmed by the high angle annular dark-field scanning transmission electron microscopy (HDDAF-STEM) (see Section I in Supplementary Information).

## Transport measurements

The resistivity of bilayer samples was measured using the standard four-point method. The superconducting transition temperatures were determined to be ∼70 and 55 K for $MnO_2$ and $La_{0.7}Ca_{0.3}O$ terminated bilayers respectively. The magnetization measurements were carried out using the superconducting quantum interference device magnetometer (SQUID, Quantum Design MPMS). 1000 Oe magnetic field was applied perpendicular to the surface of bilayers during the measurements and we only show the zero-field cooling data in the paper. The Curie temperature of LCMO layer is estimated by intersecting the linear extrapolation of high temperature leading edge of dM(T)/dT curve to zero, and is ∼190 K (see Section II in Supplementary Information).

## X-ray spectroscopy

X-ray absorption (XAS) and resonant soft x-ray scattering (RSXS) were performed at Beamline 8.0.1 of Advanced Light Source (ALS) using the RSXS endstation.[40] During the measurements, the photon energy resolution was better than 0.3 eV at Cu L-edges and the beam spot on sample was around 40 μm (v) by 500 μm (h). In XAS, the spectra shown in Fig. 2 were recorded in the total electron yield mode (photo-current from sample to the ground). The spectra were normalized by the photon flux determined from the photo-current of an upstream Au mesh. The atomic contributions responsible for step-like intensity increases around Cu $L_3$ and $L_2$ edges were further removed.

In RSXS, both incident photon polarization and scattering plane were horizontal (π-scattering geometry). A photodiode with front Al window to block out ambient light (primarily the visible light around the chamber) was used to record the scattered x-rays from the bilayers. This detector, which does not have the selectivity on the polarization or energy of scattered photons, records signal from specular reflection, fluorescence and the YBCO (001) Bragg peak simultaneously. Two types of scan, hv and q-scans, were used to produce the spectra in Fig. 2 and Fig. 3 respectively. In the momentum-space scan (q-scan), the sample and detector were placed at the specular geometry such that their angles relative to the incident photon beam followed the θ −2θ relationship. Since the signals from specular reflection and fluorescence have monotonic angular dependence, they can be separated from the Bragg peak by fitting q-scan spectra with a Lorentzian function on top of a quadratic background. The Kiessig fringes at this photon energy are negligible and do not affect the fitting results. The Bragg peak intensity presented in Fig. 3(a) -(c) and 3(f) is the peak area. The resonance profile (hv-scan) was measured with sample and detector angles following the θ −2θ relationship and tracking the incident photon energy so that the photon momentum transfer $\vec{q}$ stayed at the (001) Bragg peak wave vector.

In the azimuthal dependence measurements, the x-ray beam was defocused to a size of ∼2 mm square to mitigate the sample spatial inhomogeneity issue. The zero of azimuthal angle is defined with the crystallographic a/b axis in the horizontal scattering plane.

## DFT calculations

The first principles calculations were performed using the accurate full-potential augmented wave method, as implemented in the VASP package within the GGA+U schemes. The calculations were performed over an 8 × 8 × 1 Monkhorst-Pack k point mesh in the irreducible Brillouin zone. Both bilayers contained 4 u.c. of YBCO and 6 u.c. of LCMO. The structure of bilayer was optimized with the residual atomic forces less than 0.05 eV/ A°. U values used for La 4f and Mn 3d orbitals are 5.0 eV and 4.0 eV respectively.

**Acknowledgements**

The Advanced Light Source is supported by the Director, Office of Science, Office of Basic Energy Sciences, of the U.S. Department of Energy under Contract No. DE-AC02-05CH11231. S.W.H. would like to thank Ruimin Qiao and Wanli Yang for supporting the beamline operation. This work was also supported by MOST of Taiwan, R.O.C. under Grants 103-2112-M-009-007-MY3 and the MOE ATU program.

## Author contributions statement

S.W.H. and J.-Y.L. conceived the idea. V.T.T., J.M.L., J.M.C., Y.H.C. and J.-Y.L. provided bilayer samples and transport, magnetization characterization. S.W.H., M.C.L., Y.-D.C. and J.-Y.L. carried out the experiments. H.T.J. performed the DFT calculations. S.W.H. and S.R. performed fitting to the x-ray reflectivity measurements. S.W.H., L.A.W., Y.H.C., R.W.S., Y.-D.C. and J.-Y.L. interpreted the data. S.W.H., Y.-D.C and J.-Y.L. wrote up the manuscript.

## Additional information

**Competing financial interests** The authors declare that there is no competing financial interests.

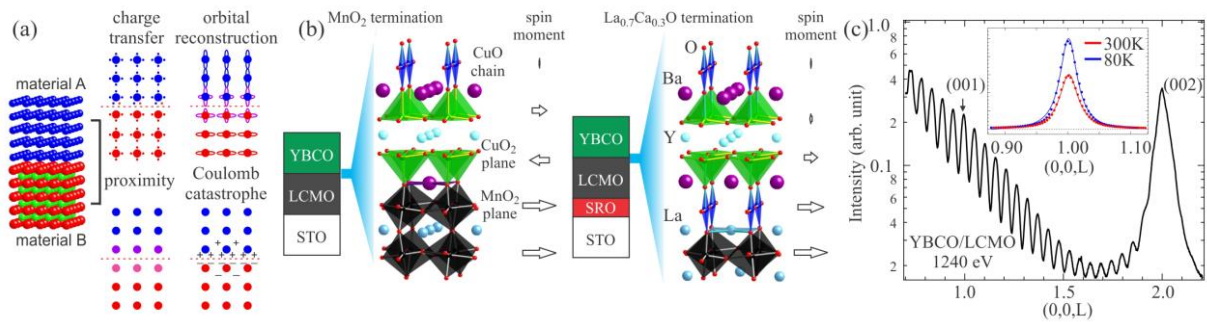

**Figure 1.** (color online) (a) Schematics illustrating various effects at the interface of a heterostructure. (b) Crystal structure near the interface of YBCO/LCMO bilayers with $MnO_2$ (left panel) and $La_{0.7}Ca_{0.3}O$ (right panel) interfacial terminations. The arrows indicate the orientation and magnitude (not in proportion) of Mn and induced Cu spin moments. (c) Synchrotron reflectivity measurement on the $MnO_2$ terminated bilayer at 80 K and 1240 eV. The YBCO (001) Bragg peak is indicated by the arrow. Inset shows the q-scans at 80 K (blue) and 300 K (red) from the $MnO_2$ terminated bilayer at 925.25 eV. The photodiode intensity was normalized by the incident photon flux only and no background subtraction was applied. The markers are data points and lines are Lorentzian fitting results.

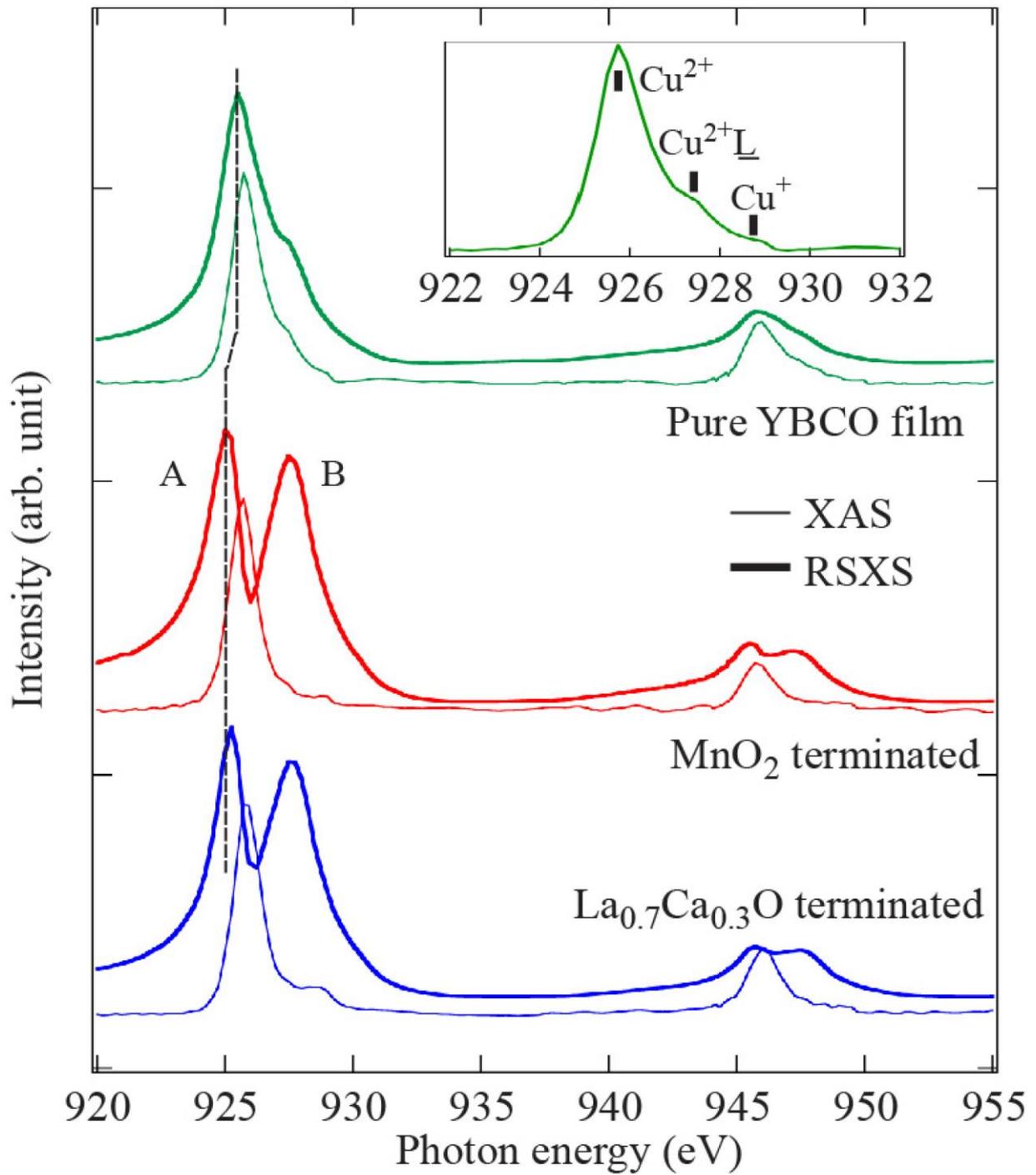

**Figure 2.** (color online) Thin and thick solid lines are the x-ray absorption spectra (XAS) and the resonance profiles of YBCO (001) Bragg peak at 300 K from the pure YBCO film (green, top panel), $MnO_2$ (red, middle panel) and $La_{0.7}Ca_{0.3}O$ terminated (blue, bottom panel) bilayers. Inset shows the Cu $L_3$ edge XAS of the pure YBCO film with ticks denoting the three states of Cu.

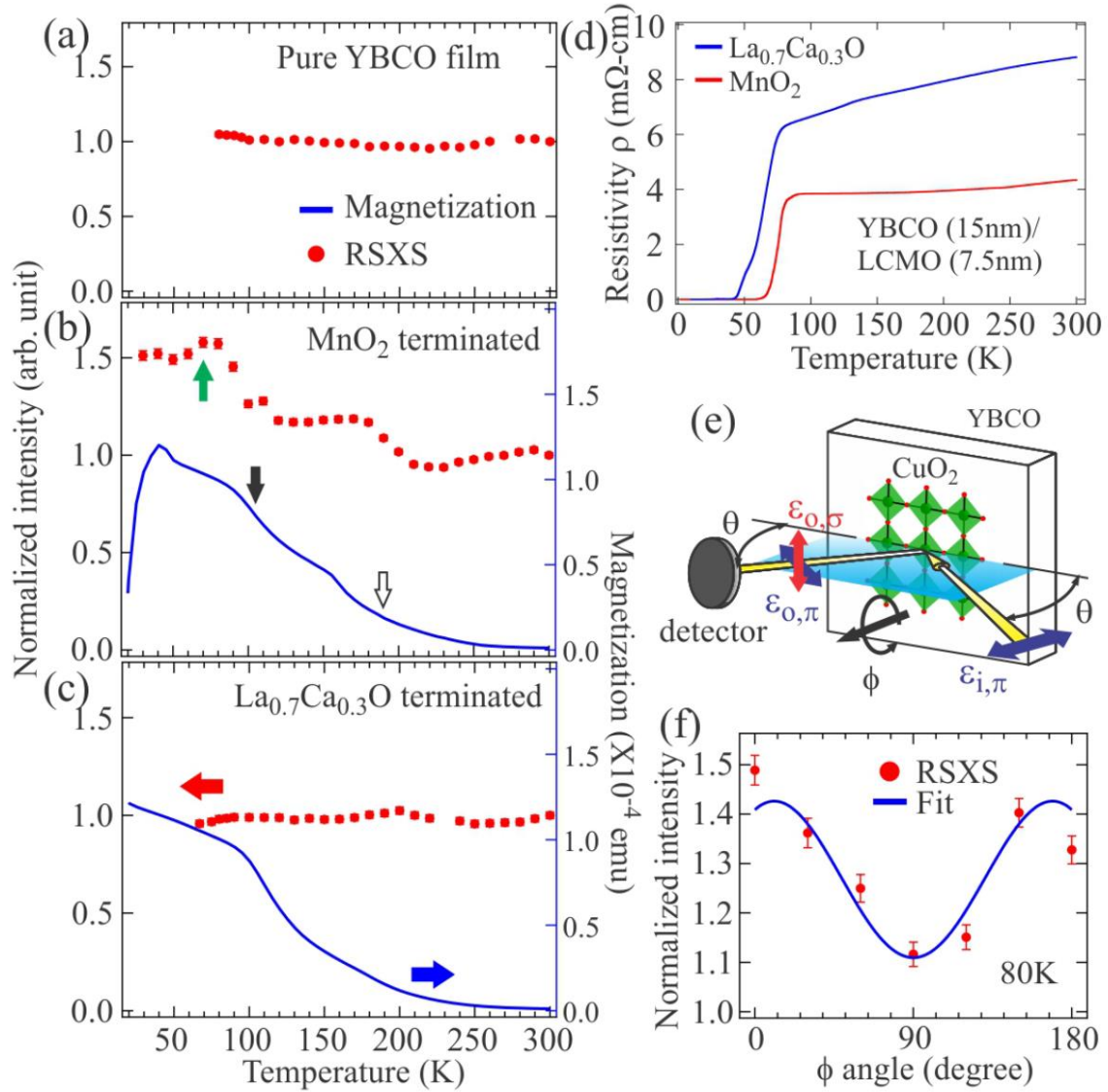

**Figure 3.** (color online) Red markers (left axis) and blue lines (right axis) represent the normalized (001) Bragg peak intensity and magnetization for (a) the pure YBCO film, (b) $MnO_2$ and (c) $La_{0.7}Ca_{0.3}O$ terminated bilayers. The intensity of (001) Bragg peak is normalized to 1.0 at 300 K. The superconducting transition, STO structural phase transition and Curie temperatures are marked by green, black and open arrows respectively. (d) Resistivity of the bilayers with MnO2 (red) and $La_{0.7}Ca_{0.3}O$ (blue) interfacial terminations. (e) Schematic illustration of experimental geometry with $\varphi$ angle at $0°$. (f) Azimuthal angle $\varphi$ dependence of the normalized (001) peak intensity (red markers) overlaid with a sinusoidal functional form (blue line, see text). The data were taken at 80 K.

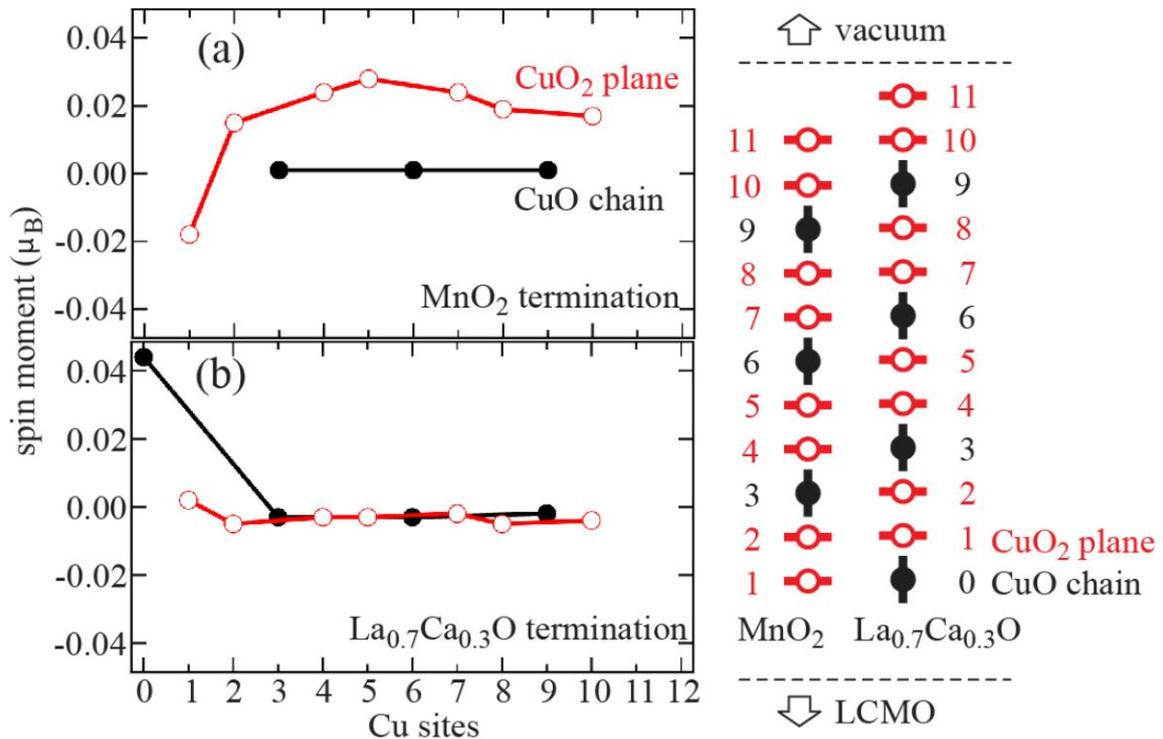

**Figure 4.** The calculated spin moment on the Cu sites in the CuO chains (black filled circles) and CuO$_2$ planes (red open circles) for (a) MnO$_2$ and (b) La$_{0.7}$Ca$_{0.3}$O terminated bilayers. The positive (negative) spin moment is defined as the Cu spin parallel (antiparallel) to the Mn spin. The schematic next to these two figures shows the tacking order of the CuO chains (black filled circles with vertical bars) and CuO$_2$ planes (red open circles with horizontal bars) along the c-axis in the calculations. The labelling of Cu sites is for the readers to associate the calculated spin moments with their spatial arrangement.

# Supplementary Information for "Selective interlayer ferromagnetic coupling between the Cu spins in $YBa_2Cu_3O_{7-x}$ grown on top of $La_{0.7}Ca_{0.3}MnO_3$"

S. W. Huang[1,2,3], L. Andrew Wray[1,4,5], Horng-Tay Jeng[6,7], V. T. Tra[8], J. M. Lee[9], M. C. Langner[2], J. M. Chen[9], S. Roy[1], Y. H. Chu[10], R. W. Schoenlein[2], Y.-D. Chuang[1,*], and J.-Y Lin[8,1,+]

[1]Advanced Light Source, Lawrence Berkeley National Laboratory, Berkeley, CA 94720, USA
[2]Materials Sciences Division, Lawrence Berkeley National Laboratory, Berkeley, CA 94720, USA
[3]MAX IV Laboratory, Lund University, P. O. Box 118, 22100 Lund, Sweden
[4]Department of Physics, New York University, New York, New York 10003, USA
[5]Stanford Institute for Materials and Energy Sciences, SLAC National Accelerator Laboratory, Menlo Park, CA 94025, USA
[6]Department of Physics, National Tsing Hua University, Hsinchu 30013, Taiwan
[7]Institute of Physics, Academia Sinica, Taipei 11529, Taiwan
[8]Institute of Physics, National Chiao Tung University, Hsinchu 30010, Taiwan
[9]National Synchrotron Radiation Research Center, Hsinchu 30076, Taiwan
[10]Department of Materials Science and Engineering, National Chiao Tung University, Hsinchu 30010, Taiwan
[*]ychuang@lbl.gov
[+]ago@cc.nctu.edu.tw

## ABSTRACT

### I. Structure of YBCO/LCMO bilayers grown by pulse laser deposition (PLD)

The ability to control the growth process down to atomic level is the key to the success of this study. In the PLD system, a 10 Hz 250 mJ pulsed laser beam from KrF (λ = 248 nm) excimer laser was used to evaporate the targets. The (100) oriented $SrTiO_3$ (STO) single crystal was used as the substrate. Before growing the bilayers, the substrate was first treated with HF-NH4F buffer solution to produce a uniform $TiO_2$ termination. Depositing $La_{0.7}Ca_{0.3}MnO_3$ on top of this $TiO_2$ terminated STO substrate produces the MnO2 termination at the $YBa_2Cu_3O_{7-x}$ - $La_{0.7}Ca_{0.3}MnO_3$ (YBCO-LCMO) interface. To produce the $La_{0.7}Ca_{0.3}O$ termination at the manganite/cuprate interface, a 1.5 unit cell (u.c.) $SrRuO_3$ buffer layer was deposited on top of the STO substrate before growing the LCMO layer. The LCMO (YBCO) layer was deposited at 700 °C (750 °C), 80 mTorr (150 mTorr) oxygen pressures. After growth, the bilayer samples were further annealed in a 550 °C, 700 Torr oxygen atmosphere for one hour followed by a slow cool down to the room temperature to achieve the full oxygenation for the YBCO layer. We used the in-situ reflection high-energy electron diffraction (RHEED) to monitor the growth process. Clear RHEED oscillations confirmed the layer-by-layer growth mode (for examples, see Fig. S1). To determine the quality of YBCO-LCMO interface, high angle annular dark-field scanning transmission electron microscopy (HDDAF-STEM) was used. The HDDAF-STEM images confirmed the structures illustrated in Fig. 1(b) in the manuscript.[1]

### II. Transport properties of YBCO/LCMO bilayers

The resistivity of bilayer samples was measured using the standard four-probe method. The results are shown in Fig. S2(a) as well as in Fig. 3(d) in the manuscript. The superconducting transition temperatures are determined to be ∼ 70 and 55 K for $MnO_2$ and $La_{0.7}Ca_{0.3}O$ terminated bilayers respectively. The Curie temperature, which is the onset temperature of ferromagnetism in the LCMO layer, is estimated by intersecting the linear extrapolation of the high temperature leading edge of dM(T)/dT curve to zero. The extrapolation is shown as the dashed line in Fig. S2(b) and the Curie temperature is estimated to be 190 K, with an uncertainly on the order of 10 K.
In Fig. 3(b) in the manuscript, the onset of RSXS intensity (red markers) is slightly higher than the Curie temperature

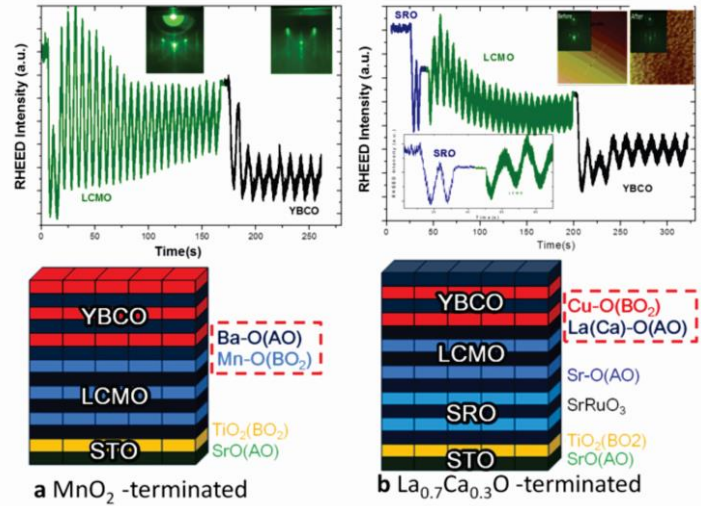

**Figure 1.** (color online) Examples of the growth of (a) the STO/LCMO$_{10nm}$/YBCO$_d$ structure with MnO$_2$-terminated interface and (b) the STO/SRO$_{1nm}$/LCMO$_{10nm}$/YBCO$_d$ structure with La$_{0.7}$Ca$_{0.3}$O terminated interface. Top panels show the layer-by-layer growth mode monitored by the RHEED.

(open arrow) and this can be understood as the following. For magnetization, the recorded signal is the sum of magnetic moments from different ferromagnetic domains. When these magnetic domains are not properly aligned, say pointing along the easy axes (Mn-O bond direction) but antiparallel to each other, cancellation can lead to a smaller reading in M(T). On the other hand, the RSXS intensity measures primarily the sum of the square of these moments. Thus even in this scenario, as long as the c-axis ferromagnetic coupling is established, the RSXS intensity from these anti-aligned domains will add up. Thus it is plausible that in the temperature regime between 200 K and 150 K, the discrepancy between M(T) and RSXS data is caused by the intricate re-alignment of microscopic ferromagnetic domains. Moreover, a short range ordering preceding the establishment of bulk ferromagnetism would also lead to a higher onset temperature for the RSXS intensity than the Curie temperature by a similar mechanism as mentioned above.

### III. Two-peak structures in the resonance profile of YBCO (001) Bragg peak

In Fig. 2 of the manuscript, we show the resonance profiles (thick solid lines) of YBCO (001) Bragg peak from MnO$_2$ and La$_{0.7}$Ca$_{0.3}$O terminated bilayers. Unlike pure YBCO film which shows just one feature at Cu L$_3$ and L$_2$ edges, bilayer samples exhibit two peaks at these edges (the additional peak at L$_3$ edge is labelled B).

The two-peak structure is intrinsic to the bilayers. Although one might speculate that it could come from two types of Cu$^{2+}$ with different binding energies, this scenario can be ruled out as the similarity between the XAS spectra from these three samples does not support two distinct Cu$^{2+}$ states with such large energy difference ($\sim$ 2.5 eV apart). Furthermore, we also try to simulate the RSXS spectra of bilayers by using the ones from pure YBCO film with a 2.5 eV relative energy shift. Although the simulated spectrum (black curve, Fig. S3) seems to capture the gross spectral lineshape, the differences can still be clearly seen at selected photon energies (see arrows in Fig. S3).

The self-absorption effect, where x-rays emitted from deep inside the sample are re-absorbed when they come out of the sample, tends to suppress the high intensity features in XAS spectra recorded in the fluorescence yield mode. This effect becomes appreciable around the elemental absorption edges at which the x-ray penetration depth is significant reduced. Self-absorption correction is often performed when the x-ray penetration depth is comparable to or shorter than the thickness of sample. In our case where the thickness of YBCO film is around 15 nm (at 35° grazing incidence angle, the effective thickness is around 26 nm) and the minimum x-ray penetration depth at Cu L$_3$ edge is around 140 nm (attenuation length determined from the CXRO website; http://henke.lbl.gov/optical constants/), the self-absorption effect is not expected to significantly alter the intensity ratio between features A and B. Thus it cannot be used to account for the observed two-peak structure in RSXS data.

The intensity of XAS is proportional to the imaginary part of atomic scattering form factor $f''(E)$, whereas the RSXS intensity is related to its square $(f'(E))^2 + (f''(E))^2$ modulated by a phase factor from the spatial arrangement of these scatterers (see description in the manuscript). $f'(E)$ and $f''(E)$ are related to each other through Kramers-Kronig relations. We notice that the intensity of YBCO (001) Bragg peak is much weaker than other (00L) Bragg peaks and this is due to an effective destructive interference between Cu charge scatterings from the CuO chain and two $CuO_2$ planes within the unit cell. Such interference can be disrupted by a slight shift in the resonance energies or variations in the spatial arrangement of the scatterers. The former one will affect the energy denominator in $f'(E)$ and $f''(E)$, whereas the later one will affect the phase factor. Simulating the RSXS lineshape will require the full knowledge of the spatial arrangement of Cu charges within the unit cell and their energetic upon heterogeneity, but to lowest order, these two factors can explain the relative intensity change between feature A and B in Fig. 2 in the manuscript. Irrespective to which origin, the distinct RSXS resonance profiles seen in the bilayer samples implies that the local energetic of electronic states is altered upon heterostructuring.

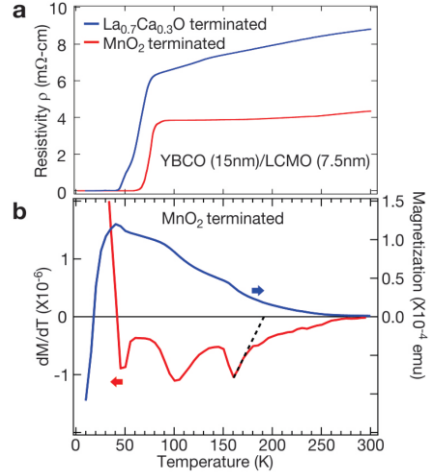

**Figure 2.** (color online) (a) The in-plane resistivity of $MnO_2$ (red curve) and $La_{0.7}Ca_{0.3}O$ (blue curve) terminated bilayers showing the respective superconducting transition temperatures $T_c$ ~70 K and 55 K determined by the mid-point of the transition. (b) Magnetization (blue curve, right axis) and its first derivative (red curve, left axis) from $MnO_2$ terminated bilayer. The dashed line indicates the linear extrapolation of the leading edge of the red curve.

## IV. Azimuthal angle dependence of the RSXS intensity

Unlike the magnetization, RSXS has the unique elemental, chemical and bonding specificity to differentiate the origins of the magnetic moments. Complementary to the x-ray magnetic circular dichroism (XMCD), RSXS intensity dependence on the tensorial nature of scattering channels can be helpful in identifying the magnetic couplings between the $CuO_2$ planes when the ordering vector overlaps with the structural Bragg peak whose intensity is dominated by the charge scattering.

In the current study, the scattering plane is horizontal and the incident photon polarization is in this scattering plane (p-polarization). We have used the single channel detector (photodiode) without polarization analyzer to record the scattering signal. The recorded signal will contain both s-and p-polarization components. Although this may complicate the analysis of spin states, we will show that it still can offer some useful insight.

We follow the formalism outlined in Hill & McMorrow[2] and Lovesey & Collins.[3] Since the incident photon energy is tuned close to $Cu^{2+}$ $L_3$ absorption edge, we only consider the dipole (E1) ransition and neglect the much weaker quadrupole (E2) transition. Equation (15) in ref[2] is reproduced here:

$$f^{(\hat{\varepsilon}_i, \hat{\varepsilon}_o)} \rightarrow \begin{pmatrix} f_{\sigma_i \rightarrow \sigma_0} & f_{\pi_i \rightarrow \sigma_0} \\ f_{\sigma_i \rightarrow \pi_0} & f_{\pi_i \rightarrow \pi_0} \end{pmatrix}$$

$$= F^{(0)}(E) \underbrace{\begin{pmatrix} 1 & 0 \\ 0 & \cos(2\theta) \end{pmatrix}}_{\hat{\varepsilon}_i \cdot \hat{\varepsilon}_o} + iF^{(1)}(E) \underbrace{\begin{pmatrix} 0 & z_1 \cos(\theta) + z_3 \sin(\theta) \\ z_3 \sin(\theta) - z_1 \cos(\theta) & -z_2 \sin(2\theta) \end{pmatrix}}_{(\hat{\varepsilon}_i \times \hat{\varepsilon}_o) \cdot \hat{s}}$$

$$+ F^{(2)}(E) \underbrace{\begin{pmatrix} z_2^2 & -z_2(z_1 \sin(\theta) - z_3 \cos(\theta)) \\ z_2(z_1 \sin(\theta) + z_3 \cos(\theta)) & -\cos^2(\theta)(z_1^2 \tan^2(\theta) + z_3^2) \end{pmatrix}}_{(\hat{\varepsilon}_i \cdot \hat{s})(\hat{\varepsilon}_o \cdot \hat{s})} \quad (1)$$

with $F^{(0)}$, $F^{(1)}$, and $F^{(2)}$ defined in ref.[2] We need to include these three terms because the Kronecker δ that conserves the wave vectors becomes 1 in this case. We only need to consider the second column in the matrix because these elements are relevant to the signal in the $\pi_i \to \sigma_0$ and $\pi_i \to \pi_0$ channels (here, $\pi_i$ and $\sigma_0$ refer to the incident $\pi$ and scattered $\sigma$ polarizations). $\theta$ is the YBCO (001) Bragg angle and is ~34.79° in the current study. $z_i$ are the components of spin unit vector projected onto three crystalline axes. They are:

$$z_1 = \sin(\alpha)\cos(\varphi)$$
$$z_2 = \sin(\alpha)\sin(\varphi) \quad (2)$$
$$z_{31} = \cos(\alpha)$$

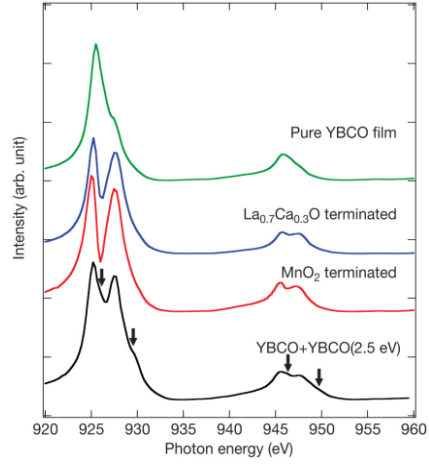

**Figure 3.** (color online) Resonance profiles of YBCO (001) Bragg peak and simulated spectrum. The black curve is produced by the sum of two spectra from pure YBCO (green curve) with a 2.5 eV relative energy shift. The discrepancies between the simulated spectrum and the ones from bilayers (red and blue curves for $MnO_2$ and $La_{0.7}Ca_{0.3}O$ terminated bilayers) are highlighted by arrows.

Here $\alpha$ is the angle between the unit vector and c-axis, and $\varphi$ is the sample azimuthal angle. Firstly, it is clear that if the moment is completely along the c-axis ($\alpha = 0°$), there will be no azimuthal angle dependence in the RSXS intensity. To simplify the discussion, we consider the extreme case where $\alpha = 90°$. Putting these terms together, we have:

$$f \sim \begin{pmatrix} f_{\sigma_i \to \sigma_0} & -i(z_1\cos(\theta))F^{(1)} - z_2(z_1\sin(\theta))F^{(2)} \\ f_{\sigma_i \to \pi_0} & F^{(0)}\cos(2\theta) + iz_2z_1\sin(2\theta)F^{(1)} - z_1^2(\sin^2(\theta))F^{(2)} \end{pmatrix} \quad (3)$$

The RSXS intensity is proportional to sum (over the superlattice) of the square of f

$$|f|^2 = |F^{(0)}|^2 \Big\{ \cos^2(2\theta) - \frac{2\Im(F^{(0)} * F^{(1)})}{|F^{(0)}|^2}\cos(2\theta)\sin(2\theta)\sin(\varphi) - \frac{2\Re(F^{(0)} * F^{(2)})}{|F^{(0)}|^2}\cos(2\theta)\sin^2(\theta)\cos^2(\varphi)$$

$$+ \frac{2\Im(F^{(2)} * F^{(1)})}{|F^{(0)}|^2}(2\sin^2(\theta) - 1)\sin(\theta)\cos(\theta)\cos^2(\varphi)\sin(\varphi)$$

$$+ \frac{|F^{(1)}|^2}{|F^{(0)}|^2}(\cos^2(\theta)\cos^2(\varphi) + \sin^2(2\theta)\sin^2(\varphi))$$

$$+ \frac{|F^{(2)}|^2}{|F^{(0)}|^2}(\sin^4(\theta)\cos^4(\varphi) + \sin^2(\theta)\sin^2(\varphi)\cos^2(\varphi)) \Big\} \quad (4)$$

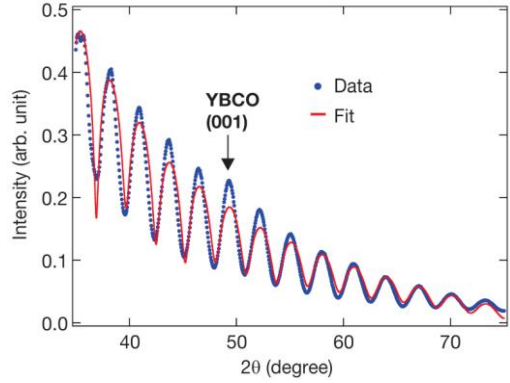

**Figure 4.** (color online) Synchrotron reflectivity measurement. The agreement between data (blue markers) and fit (red line) implies that the choice of 15 nm YBCO and 7.5 nm LCMO layer thickness with 0.6 nm roughness in fitting is reasonable. The discrepancy around 50° is due to the presence of YBCO (001) Bragg peak.

Usually, $F^{(0)}$ is much larger than $F^{(2)}$ and $F^{(1)}$ so that the ferromagnetic signal would be very weak compared to the charge signal in the Bragg peak. However, the destructive interference between charge scatterings leads to a much weaker YBCO (001) Bragg peak (see previous discussion). This makes the ratio $\sum_{\text{superlattice}}[F^{(1,2)}/F^{(0)}]$ not so negligible. But even so, we do not expect the ratio can be on the order of 1. Thus we argue that the high order terms (last three terms in the equation) can be dropped out in the following discussion. By doing so, the scattering intensity will have the azimuthal angle dependence of $a + b*\sin(\varphi) + c*cos^2(\varphi)$ where the coefficients b and c are related to $\frac{F^{(1)}}{F^{(0)}}$ and $\frac{F^{(2)}}{F^{(0)}}$.

From Fig. 3(f) of the manuscript, the Bragg peak intensity changes from ∼1.5 at 0° to ∼1.1 at 90° above the charge background of ∼1.0. Having the spin moment along the c-axis would increase the constant base line and reduce the $[F^{(1,2)}/F^{(0)}]$ ratio. The strong sinusoidal oscillation implies that the in-plane spin component is larger than the out-of-plane component. It also suggests that $F^{(1)}$ is much smaller than $F^{(2)}$, as expected from the extremely weak Cu XMCD versus XLD signal.

Although YBCO has CuO chains that naturally break the four-fold symmetry, the bilayer samples under study are twinned. It is possible that the twinned domains with two distinct CuO chain orientations have unequal volume fractions that give the observed two-fold symmetry, we also notice that the 80 K measurement temperature is below the structural distortion of underlying STO substrate around 105 K.[4, 5] This distortion naturally breaks the four-fold symmetry and further aligns the ferromagnetism in LCMO layer as shown in the magnetization.

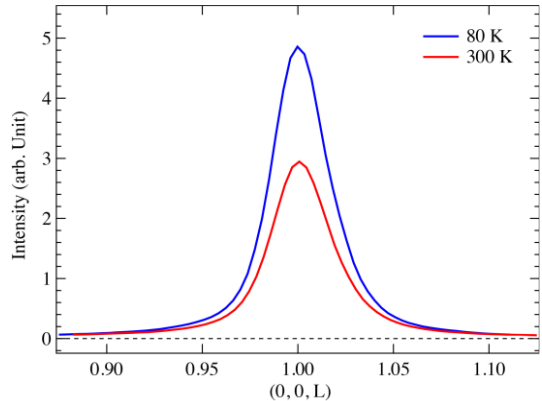

**Figure 5.** (color online) q-scans at 80 K (blue) and 300 K (red) from the $MnO_2$ terminated bilayer at 925.25 eV. The photodiode intensity was normalized by the incident photon flux only (photocurrent from upstream Au mesh), and no background subtraction was applied.

## V. Determining the layer thickness using synchrotron reflectivity

Synchrotron reflectivity is used to determine the YBCO and LCMO layer thickness in the heterostructures. The measurement temperature was set to 80 K and the incident photon energy was tuned to 1240 eV, well above the Mn and Cu resonances. The blue markers in Fig. S4 are the data while the red curve is the fitting with YBCO / LCMO layer thickness of 15 nm / 7.5 nm respectively with roughness around 0.6 nm. The agreement between data and fit justifies the fitting parameters of layer thickness and roughness. Note that the discrepancy around 2θ ∼50º is caused by the YBCO (001) Bragg peak.

The Kiessig fringes that overwhelm the YBCO (001) Bragg peak become negligible when the incident photon energy is tuned to Cu $L_3$ resonance. As can be seen in Fig. S5, the spectra recorded in q-scan can be nicely fitted by a Lorentzian function on top of a monotonic background.